\documentstyle[12pt,psfig,fleqn]{article}
\begin{document}
\vspace{2cm}
{\centering\bf DOES NEUTRINO REALLY EXIST ?\\}
\bigskip
{\centering Yu.A. BAUROV\\}
\vskip10pt
{\centering \it Central Research Institute of Machinery\\}
{\centering \it 141070, Pionerskaya 4, Korolyov,  Moscow Region, Russia\\}
\bigskip
\abstract{
An analysis of known experiments on direct and indirect
detection of the electron-type neutrino and antineutrino has
been made. The analysis based on a new hypothesis that the
observed physical space is formed from a finite set of {\it byuons},
"one-dimensional vectorial objects".  It is shown in the article
that the hypothesis for existence of neutrinos advanced by Pauli
on the basis of an analysis of the conservation laws, is not
unquestionable since the fulfillment of these laws may be secured
by the physical space itself (physical vacuum) being the lowest
energy state of a discrete oscillating system originating in the
course of byuon interaction.  This effect is analogous to that
of M\"ossbauer.  The direct experiments on detecting neutrinos
are explained from the existence of a new information channel
due to the uncertainty interval for coordinate of the
four-contact byuon interaction forming the interior geometry of
elementary particles and their properties.  Given are also the
results of an experiment on observation of cyclic variations of
the $\beta$-decay rate, which confirm the existence of said new
information channel.}

\pagebreak

\section{Introduction.}

As is known \cite{1}, neutrino (electron-type one) was discovered by
Pauli "on the tip of his pen", on the basis of analysis of
conservation laws. Recently, a number of papers \cite{2}-\cite{10} has
evolved which extends considerably our knowledge of neutrino
physics.  In Refs. \cite{2}-\cite{7}, it is shown by various research teams
on the basis of investigation of tritium spectrum end that mass
of neutrino is less than zero.  In Ref. \cite{8}, results of
precision measurements of tritium spectrum end are presented
which suggest that mass squared of the electron-type neutrino
$m^2_{\nu_e} < 0$. This artifact is explained in Ref. \cite{9} with the aid of
long range interactions of anomalous neutrinos through
introducing a potential equal to the sum of those of  Yukawa's
type.  In Ref. \cite{10}, these unique experimental results are
explained by introducing a weakly interacting light scalar boson
through which the neutrinos interact in some cloud. It is shown
in so doing that the parameters of neutrinos will be cloud
density dependent. This work develops the assumption of Ref. \cite{2}
that to account for the experiments, the density of neutrinos is
necessary which is $10^{13}$ times greater than that accepted in
cosmology.

In the new physical conception \cite{11,12} of forming
the observed three-dimensional physical space ${\bf R_3}$ from a finite
set of one-dimensional objects ({\it byuons}) in consequence of their
dynamics in a one-dimensional space ${\bf R_1}$, the most simple object,
under which the author means the electron-type neutrino, has
just the imaginary self rest energy $m_{\nu_e} c^2$  in ${\bf R_1}$ and the
imaginary quantum number, mass $m_{\nu_e}$, in ${\bf R_3}$. That is why the
question in the heading of this article has arisen: "Does the
neutrino really exist?" In order to understand what is the
matter and to be acquainted with the terminology of the subject,
let us present fundamental concepts and conclusions of the
theory \cite{11,12}.

\section{Fundamental~Theoretical~Concepts~of 
Physical Space Origin.}

In Ref. \cite{11}, the conception of formation of the observed
physical space ${\bf R_3}$ from a finite set of one-dimensional discrete
"magnetic" fluxes, new physical objects which will henceforward
be named {\it byuons} \footnote{in connection with the numerous comments:
in order to make distinction between these new physical objects and the
conventional magnetic fluxes existing in classical electrodynamics.}
, was firstly formulated. In quantum field
theory, the space observed is usually given \cite{13}, and in modern
superstring and supersymmetric theories, this space (Minkovsky's
one) is obtained through compactification of "excessive"
dimensions \cite{14,15,16}.  The {\it byuons} ${\bf \Phi}(i)$ are one-dimensional vectors
and have the form: $$ {\bf\Phi}(i) = {\bf A_g} x(i),$$ where $x(i)$ is the "length" of the
byuon, a real (positive, or negative) value depending on the
index $i = 0,1,2,...,k....$, a quantum number of ${\bf \Phi}(i)$; under $x(i)$ a
certain time charge of the byuon may be meant \cite{17}.  The vector
${\bf A_g}$ represents the cosmological vectorial potential, a new basic
vectorial constant \cite{11,12,18}. It takes only two values:
${\bf A_g} = \{A_g , -\sqrt{-1}A_g\}$,
where $A_g$  is the modulus of the cosmological vectorial potential
($|{\bf A_g}| = 1.95\cdot 10^{11}$ CGSE units).

According to the theory \cite{11, 12}, the value ${\bf A_g}$ is the limiting one.
In reality, there exists in nature, in the vicinity of the Earth, a certain summary
potential ${\bf A_{\Sigma}}$, i.e. the vectorial potential fields from the Sun
($A_{\odot} \approx 10^8$ CGSE units), the Galaxy ($A_{G} \approx 10^{11}$ CGSE units), and the
Metagalaxy ($A_{M} > 10^{11}$ CGSE units) are superimposed on the constant ${\bf A_g}$
resulting probably in some turning of ${\bf A_{\Sigma}}$ relative to the vector
${\bf A_g}$ in the space ${\bf R_3}$ or in a decrease of it.

Hence in the theory of physical space (vacuum) which the present article leans upon,
the field of the vectorial potential introduced even by Maxwell
gains a fundamental character. As is known, this field was
believed as an abstraction. All the existing theories are
usually gauge invariant, i.e., for example, in classical and
quantum electrodynamics, the vectorial potential A is defined
with an accuracy of an arbitrary function gradient, and the
scalar one is with that of time derivative of the same function,
and one takes only the fields of derivatives of these
potentials, i. e. magnetic flux density and electric field
strength, as real.  ln Refs. \cite{11,12,18,19}, local violation of
the gauge invariance was supposed, and the elementary particle
charge and quantum number formation processes were investigated
in some set, therefore the potentials gained an unambiguous
physical meaning there. In the present paper, this is a finite
set of {\it byuons}.  The works by D.Bohm and Ya.Aharonov \cite{20}-\cite{23}
discussing the special meaning of potentials in quantum
mechanics are the most close to the approach under
consideration, they are confirmed by numerous experiments
\cite{24,25}.  The {\it byuons} may be in four  vacuum states (VS)
$II^+,I^+,II^-,I^-,$ in which they discretely change the value ot
$x(i)$:  the state $II^+$ discretely increases and $I^+$ decreases
$x(i) > 0$, the states $II^-$ and $I^-$ discretely increase or decrease
the modulus of $x(i) < 0$ , respectively.  The sequence of discrete
changes of $x(i)$ value is defined as a proper discrete time of
the byuon. The byuon vacuum states originate randomly \cite{11,12}.
In Refs. \cite{11,12}, the following hypothesis I has been put
forward.  It is suggested that the space ${\bf R_3}$ observed is built up
as a result of minimizing the potential energy of byuon
interactions in the one-dimensional space ${\bf R_1}$ formed by them.
More precisely, the space ${\bf R_3}$ is fixed by us as a result of
dynamics arisen of {\it byuons}. The dynamic processes and, as a
consequence, wave properties of elementary particles appear
therewith in the space ${\bf R_3}$ for objects with positive potential
energy of byuon interaction (objects observed).  Let us briefly
list the results obtained earlier  when investigating the
present model of physical vacuum:

2.1. The existence of a new
long-range interaction in nature, arising when acting on
physical vacuum by the vectorial potential of high-current
systems, has been predicted \cite{26}-\cite{30}.

2.2. All the existing
interactions (strong, weak, electromagnetic and gravitational
ones) along with the new interaction predicted have been
qualitatively explained in the unified context of changing in
three periods of byuon interactions with characteristic scales -
$10^{-17}$ cm, $10^{-13}$ cm, and $10^{28}$ cm, determined from the minimum
potential energy of byuon interaction \cite{12,17}.

2.3. Masses of leptons, basic barions and mesons have been found \cite{18, 33}.

2.4. The constants of weak interaction (vectorial and axial
ones) and of strong interaction have been calculated \cite{18, 33}.

2.5. The origin of the galactic and intergalactic magnetic
fields has been explained as a result of existence of an
insignificant ($\approx 10^{-15}$) asymmetry in the formation process of ${\bf R_3}$
from the one-dimensional space of {\it byuons} \cite{11,12}.

2.6. The substance density observed in the Universe ($10^{-29} g/cm^3$ ) has
been calculated \cite{11,12}.

2.7. The origin of the relic radiation has been explained on the basis of unified mechanism of the
space ${\bf R_3}$ formation from one-dimensional space ${\bf R_1}$ of {\it byuons}
\cite{11,12}, etc.

Let us explain item 2.1 briefly. It is shown in Ref. \cite{33} that masses of all elementary particles are
proportional to the modulus of ${\bf A_g}$ . If now we direct the
vectorial potential of a magnetic system in some space region
towards the vector ${\bf A_g}$ then any material body will be forced out
of the region where $|{\bf A_\Sigma}| < |{\bf A_g}|$ . The new force is nonlocal,
nonlinear, and represented by a complex series in $\Delta A$, a
difference in changes of $|{\bf A_g}|$ due to the potential of a current
at location points of a sensor and a test body \cite{12,29}. This
force is directed mainly along the vector ${\bf A_g}$, but as the latest
experiments have shown, there is also an isotropic component of
the new force in natures, which component acts
omnidirectionally from the space of the maximum decrease in
$|{\bf A_g}|$.  Corresponding to it are probably the even terms in a
series representing the new force \cite{12,29}.

One of the important predictions of the theory is revealing a new information
channel in the Universe which is associated with the existence of a
minimum object with positive potential energy, so called object
{\bf 4B}, arising in the minimum four-contact interaction of {\it byuons} in
the vacuum states $II^+,I^+,II^-,I^-$. In four-contact byuon
interaction, a minimum action equal to $h$ (Planck's constant)
occurs, and the spin of the object appears. Hence the greater
part of the potential energy of byuon interaction is transformed
into spin of the object {\bf 4B}. The residual (after minimization)
potential energy of the object {\bf 4B} is equal to $\approx 33 eV$, it is
identified with the rest mass of this object in the space ${\bf R_1}$.
In agreement with Refs. \cite{11,12}, the indicated minimum object
{\bf 4B} has, according to Heisenberg uncertainty relation, the
uncertainty in coordinate $\Delta x \approx 10^{28} cm$ in ${\bf R_3}$.
The total energy of these objects determines near $100 \%$ energy of the Universe as
well as its substance density observed.  In such a manner the
objects {\bf 4B} connect together, due to the uncertainty relation,
all the elementary particles of the Universe and hence all the
objects in the animate and inanimate nature as composed of
elementary particles.  The greater is the number of elementary
particles of substance in some place of space, the greater is
also the number of objects {\bf 4B} there because the latters form the
interior geometrical space of elementary particles \cite{11,12}.

\section{Analysis of experiments on detecting neutrino.}

All the experiments proving the existence of  neutrino may be
divided into two groups in the first of which we are dealing
with the circumstantial evidence, starting from the conservation
laws, that neutrinos exist \cite{34,35} among these are also the
papers \cite{2}-\cite{8}, whereas the second group of experiments has
pretensions to the direct corroboration of neutrino existence
\cite{36, 37}.

Let us consider the first experiment on detecting
neutrino, which was carried out by A.l.Leipunsky \cite{36} and
relates to the first group of above mentioned ones. The idea of
his experiment was constructed on comparison of energetic
spectra of electrons and recoil nuclei produced during the
$\beta$-decay. If a neutrino (antineutrino) were not emitted in this
process, the law of conservation of momentum would be obeyed:
$$ {\bf P_e} + {\bf P_{r,n}} = 0, \hspace{1cm} |P_e| = |P_{r,n}|\eqno{(1)}$$
where $P_e$  is momentum of  $\beta$-electron, $P_{r,n}$ is momentum obtained
by the recoil nucleus during $\beta$-decay.  If  however a neutrino
(antineutrino) is emitted in  $\beta$-decay, the law of momentum
conservation has the form
$$ {\bf P_e} + {\bf P_{r,n}} + {\bf P_{\nu}} = 0, \eqno{(2)}$$
and then
$$ |P_e| \ne |P_{r,n}|.\eqno{(3)}$$
A.l.Leipunsky, when investigating the process of $\beta^+$-decay in his experiments with
the carbon isotope ${}^4_6C$ has validated the inequality (3) and
thereby, as he believed, proved the existence of neutrino.  The
Leipunsky's experiment and those in the first group can be
explained on the assumption that the conservation laws of
momentum, angular momentum etc., are taken over by a large-scale
object, the physical space (physical vacuum) of the Universe.
An analogy to such a phenomenon is the known {\it M\"ossbauer effect}
\cite{38} lying in the fact that the resonant absorption of $\gamma$-radiation by nuclei in conditions of partial overlap of emitted
and absorbed $\gamma$-radiation lines rises sharply when cooling the
radiation source and absorbent. M\"ossbauer had accounted for
this strange (for the year 1958) effect by the fact that, in
certain situations (sufficiently low transition energy, low
temperature as compared with the Debye temperature of the
crystal), the recoil momentum and energy produced during
emission (absorption) of  $\gamma$-quantum do not go into either
knocking out an atom from a site of the lattice or changing the
energy state of the crystal but are transmitted, in elastic
manner, to the entire crystal or at least to a large group of
atoms embraced by travelling acoustic wave during the emission
time. In such a case, the correlation between the momentum and
energy of the emitting (absorbing) nucleus breaks down since the
recoil energy practically equals zero due to large mass of the
crystal, and hence the  $\gamma$-quantum energy difference between
the emission and absorption lines practically disappears.  The
essence of the M\"ossbauer effect is that an oscillator being in
the state with minimum energy can, within the framework of
quantum model of solid body \footnote{The solid body is a set of harmonic
oscillators \cite{38}}, solely acquire energy but cannot
give it up.

An analogous pattern is observed also in the
experiment of A.l.Leipunsky as well as in other ones, for
instance, in decay of neutron :
$$ n \to p + e^- + \tilde \nu_e \eqno(4) $$
if assumed that the physical space is a certain periodic oscillating medium
taking over the laws of conservation, i.e. one may not introduce into
Expr. (4) the electron-type antineutrino as it was done by Pauli
in order to meet the conservation laws, but therewith the event
(4) may not already be considered as a local phenomenon.

Let us show that on the theoretical basis \cite{11, 12}.

The vacuum states $II^+,I^+,I^-,II^-$ of the {\it byuons} appear randomly and are
characterized for the byuon $\Phi (i+1) = {\bf A_g} x(i+1)$ by special functions
$\Psi^{i+2}_{II^+},\Psi^i_{I^+},\Psi^{k-i-2}_{I^-}$ and $\Psi^{k-i}_{II^-}$,
respectively. Under the product $\Psi^{i+2}_{II^+}\Psi^i_{I^+}$ is meant,
in Refs. \cite{11,12}, the probability of  existence of  vacuum
states $II^+$  and $I^+$  of {\it byuons} with the index $(i+1)$. Recall that
the vacuum state $II^+$ increases the index $i$ of a byuon, i.e. the
value $x(i+1)$, by one, and the vacuum state $I^+$ decreases it by one.
Normalizing the functions $\Psi_{II^+}$, $\Psi_{I^+}$, $\Psi_{I^-}$ ,
$\Psi_{II^-}$, for the case $i \le k$ gives \cite{11,12}:
$$ \sum\limits^{(NkP-k)/2}_{\xi = 0}\sum\limits^{j = i}_{j = 0}
\Psi^{j+2}_{II^+},\Psi^j_{I^+},\Psi^{NkP-j-2\xi}_{II^-}\Psi^{NkP-j-2-2\xi}_{I^-} = \frac{NP}{2}\eqno{(5.1)}$$
$$ \sum\limits^{(NP-1)}_{\xi = 1}\sum\limits^{j = i}_{j = 0}
\Psi^{NkP-j}_{II^-}\Psi^{NkP-j-k\xi}_{II^-} = P\eqno{(5.2)}$$
$$ \sum\limits^{(NkP-k)/2}_{\xi = 0}\sum\limits^{j = i}_{j = 0}
\Psi^{j+2}_{II^+}\Psi^{NkP-j-2-2\xi}_{I^-} = \frac{NP}2\eqno{(5.3)}$$
$$ \sum\limits^{(NkP-k)/2}_{\xi = 0}\sum\limits^{j = i}_{j = 0}
\Psi^{j}_{I^+}\Psi^{NkP-j-2\xi}_{II^-} = \frac{NP}2\eqno{(5.4)}$$
Here N,k,P  are periods of {\it byuon} interaction
in i, they are computed precisely in Refs.\cite{12,17} based on the
minimum potential energy of byuon interaction in ${\bf R_1}$ (N,k, and P
are integer numbers equal to $1.54\cdot 10^4$,$3.2\cdot 10^{15}$ ,$10^{42}$,
respectively). To these periods there correspond the distances
$k\tilde x_0 \approx 3\cdot 10^{-17} $cm;$Nk\tilde x_0 \approx 10^{-13}$cm;
$NkP\tilde x_0 \approx 10^{28}$ cm, where $\tilde x_0 \approx 2.8\cdot 10^{-33}$cm
is a quantum  of  space such as $\tilde x_0 / \tau_0 = c $ ($c$  is speed of
light, $\tau_0 \approx 0,9\cdot 10^{-43} s$  is a time quantum).

It turned out to be remarkable and surprising that, for the four-contact
interaction of {\it byuons} ({\bf 4B}-objects) when at a single discrete point of the
discrete one-dimensional space ${\bf R_1}$ {\it byuons} can be observed at a
time (i.e, in quantum of time $\tau_0$) in four vacuum states
$II^+,I^+,I^-,II^-$, the equation for the $\Psi$-function has, for those
states, the form of that of  harmonic oscillator \cite{12}:
$$\Delta(\Psi^{i+1}_{II^+} + \Psi^{i+1}_{I^+}) +
(\Psi^{i+1}_{II^+} + \Psi^{i+1}_{I^+}) = 0,\eqno{(6.1)}$$
$$\Delta(\Psi^{NkP-i-1}_{II^-} + \Psi^{NkP-i-1}_{I^-}) +
(\Psi^{NkP-i-1}_{II^-} + \Psi^{NkP-i-1}_{I^-}) = 0,\eqno{(6.2)}$$
Here $\Delta$ denotes second finite differences in index i.
It is seen from Eqs. (5) that the object {\bf 4B} is determined with probability 1 on
the characteristic dimension in ${\bf R_3}$ equal to $\approx 10^{-17}$ cm (the
completion of forming ${\bf R_3}$).  It was shown in Ref. \cite{18} that the
charge and quantum numbers of elementary particles are formed on
distances of $10^{-17}\div 10^{-13}$ cm, and to the free object {\bf 4B}
corresponds in ${\bf R_3}$, in accordance with the Heisenberg uncertainty
relation, an uncertainty in coordinate equal to $10^{28}$cm \cite{11,12}.
The residual, from minimization process, potential energy in ${\bf R_1}$
is considered in Refs. \cite{11,12} as a rest energy $mc^2$ in ${\bf R_1}$. The
energy of the object {\bf 4B} equivalent to $mc^2$ is equal to ${E^0_k}_{min}
\approx 33 eV$. This object is a boson with spin 1, there corresponds a
pair of electron-type neutrino-antineutrino ($\nu_e \Leftrightarrow \tilde\nu_e$)
 to it. The neutrino is a spinor produced by the interaction of {\it byuons} in VS
$II^+ I^+$, but its $mc^2$ is imaginary.  The smearness of objects {\bf 4B}
over the Universe gives the observed density of substance
$\approx 10^{-29} g/cm^{-3}$ and connects all the objects of the Universe into a
single information field.

Let us return to reactions of  (4)-type.
  Thus it is stated that the momentum, energy, spin,
and lepton charge attributed to the antineutrino are given up,
when the reaction (4) proceeds, to the physical space formed as
a result of minimizing the potential energy  of {\it byuons} in ${\bf R_1}$ and
observed by us as an assemblage of the objects {\bf 4B} which is
developing constantly due to the vacuum state $II^+$ and creating
an oscillating system being in the lowest energy level. Well,
but what of the experiments with direct detection of neutrino
which particle is observed repeatedly in leading nuclear
laboratories in the world (for example, the known experiment of
Cowen and Rayness \cite{34,35}):
$$ \tilde \nu_e + p \to n + e^+ \eqno(7) $$
An answer is simple. The nuclear reactors near which the reaction (7) is observed due to the
reaction (4), create around themselves an unobservable field of
objects \footnote{The self energy is imaginary.}
produced during interaction of {\it byuons} in the vacuum
states $I^-II^-$ (antineutrino), which objects in turn, when
connected with the objects created by {\it byuons} in the vacuum
states $I^+II^+$, give bosons being in the vacuum states $I^+II^+$
$I^-II^-$. The assemblage of these bosons forms the physical space,
or the space of the elementary particles. The laws of
conservation are therewith met to a high accuracy, however not
in a local form but in a volume with a characteristic dimension
 of $\approx 10^5$ cm.  Let us show this.

According to the basic hypothesis (1) the objects {\bf 4B} forms,
as said above, the interior space of an elementary particle along with all its quantum
numbers and charges. Therefore this object always creates, due
to its perpetual dynamics in the space ${\bf R_3}$, the minimum momentum
for an elementary particle as an integer entity whose interior
geometrical space is formed by it.  The momentum of the object
{\bf 4B} corresponding to the minimum one for elementary particles,
may be represented in a general form \cite{11,12,39}
$$ p = \Phi {E^0_k}_{min} / c $$
where $\Phi$ is probability of detecting the object {\bf 4B} in some region
of the space ${\bf R_3}$.  If  the objects {\bf 4B} are free (i.e. not an
elementary particle but a space free of  elementary particles is
created by them), then $$\Phi = \frac{1}{16}\frac{\tilde x_0^3}{4\pi x_0^2 \tilde x_0},$$
where $ x_0\approx 10^{-17}$ cm \cite{11,12}.
In this case, if the spread in values of momentum for an
elementary object $\Delta p$ is set equal $P$, the uncertainty in
coordinate in ${\bf R_3}$ for the object {\bf 4B} will comprise $10^{28}$ cm.
 The coefficient $1/16$ in the formula for $\Psi$ is determined from the
combinatorics of the {\it byuons} in the vacuum states $I^+,II^+, I^-,II^-$.
If the object {\bf 4B} is not free (i.e. it forms the interior
geometry of an electron, as an example), then
$$\Phi = \frac{1}{16}\frac{ x_0^3}{4\pi {(N x_0)}^2 x_0}\eqno(8)$$
and for an assemblage of $N$ objects {\bf 4B} forming an electron
(for which $m_e c^2 = N {E^0_k}_{min}$)
we may write
$$\Delta p = \frac{1}{16}\frac{\tilde x_0^3}{4\pi {(N x_0)}^2 \tilde x_0}
\frac{N {E^0_k}_{min}}{c} = \frac{1}{64\pi}\frac{{E^0_k}_{min}}{N c}.\eqno(9)$$
When using Eq. (9), we have, for $N$ objects {\bf 4B},
an uncertainty in coordinate $\Delta x$ on the order of 10 cm in ${\bf R_3}$, i.e.
an electron, due to wave properties of $N$ objects {\bf 4B}, carries
information on its properties not over distances of $10^{-8}$ cm (de Broglie
wave at the temperature $\sim 300 $K) as in the case if it were
considered as a pointwise particle, but at distances of the
order of 10 cm.  If one considers not $N$ objects {\bf 4B} in an
electron but only one object {\bf 4B} (however in an electron, i.e. $\Phi$
is determined by formula (8)) then $\Delta x \approx 10^5$ cm. Thus a lesser
quantity of information on the state of interior space
characteristic of an electron has a greater spread in
coordinate.  Hence in the range of uncertainty in coordinate
$\sim 10^5$ cm around an electron and consequently a neutron, the
processes associated with transmitting energy, momentum etc, to
the free space as well as to other objects in accordance with
the reaction (4) and (7) can occur. As is seen, the overlapping
of the processes (4) and (7) is tremendous if the reaction (7)
is detected even at a distance of hundreds meters from the
reactor, which, let us underline it once again, is in perfect
analogy to the resonant absorption of $\gamma$-radiation by nuclei and
to the M\"ossbauer's  effect as well, however not in coordinates
$\Delta E, \Delta t,$ but $\Delta p, \Delta x$. One can do therewith a simple estimation of
the cross-section of the reaction (7) using our approach if to
represent it in the form $\sigma = 1/n\lambda$ where $n$ is concentration of
nucleons in a nucleus, and $\lambda$ is the maximum uncertainty interval
for the new information field of the nucleon which field makes
possible the intersection of the reactions (4) and (7). When put
$n = 10^{38} cm^{-3}$  and $\lambda \approx 10^5 cm$, as said above, then we have
$\sigma \approx 10^{-43} cm^2$. As is known, just such a value of $\sigma$  is observed in
the experiment with neutrinos from a reactor as well as found on
the basis of standard phenomenological theories \cite{38}.

\section{Experiment  on  influence  of vectorial potential  on
$\beta$-decay rate.}

Data for influence of changes in the physical space structure,
and more precisely, in $A_\Sigma$ depending on the vectorial potentials
of the Sun and Earth, on reactions like that in Expr. (4), are
given in Ref. \cite{40}. In Fig.I is shown the time variation of
decay number for Cs-137 preparation during runs of measurements
of Apr.19-23, 1994. The experiment was based on the assumption
that probability W of $\beta$-decay is proportional to $A_g$\cite{33}.
Hence, if we diminish $A_g$ by means of a vectorial potential, the
Earth's one for example, the decreasing in $\beta$- decay numbers must
be observed. In Fig.1, the pronounced 24-hour cycles are seen
(the deflection is equal to $6\sigma$ where $\sigma \approx 4307$ is standard
deviation). The minimum number of decays corresponds therewith
to 8-10 hours of Moscow time when the point of maximum decrease
in the modulus of ${\bf A_g}$ due to the Earth's vectorial potential $A_E$
goes through the meridian of Sanct-Petersbourg (Russia) where
the experiment was carried out, under the condition that the
coordinates of the vector ${\bf A_g}$  are: the right ascension
$\alpha \approx 270^\circ \pm 7$, declination $\delta \approx 34^\circ$
\cite{26}-\cite{31} (the second equatorial coordinate
system). In Appendix are described the statement, procedure,
and equipment of the experiment.  The results of the experiment
carried out confirm that the physical space is a complex
formation, and its "breathing" tells, through the potentials, on
such a fundamental phenomenon as $\beta$-decay which is influenced on,
according to the existing standard physics, practically by no
factors (pressure, magnetic and electric fields etc.) \cite{41,42}.
Thus, according to the conception being developed on formation
of physical space from a finite set of {\it byuons}, the invoking the
Pauli's hypothesis on the existence of neutrino is by no means
necessary to explanation of  weak interactions.

\vskip15pt
{\centering \large \bf APPENDIX \\}
\vskip15pt

In Ref. \cite{18}, the expressions for the vectorial constant $C_V$ and
the axial one $C_A$ of weak interaction in terms of ${\bf A_g}$, elementary
electric charge $e$, and periods of byuon interaction, are found.
These periods are determined in Refs. \cite{12,17} mathematically
from the minimum potential energy of byuon interaction in the
one-dimensional space formed by them ($x_0 \approx 10^{-17}$ cm is the period
characterizing the completion of ${\bf R_3}$ formation; $N x_0 = ct^* \approx 10^{-13}$
cm characterizes formation of elementary particle electric
charge in space \cite{12,18}):
$$|C_V| = e |{\bf A_g}|\frac{2 x_0}{c t^*}\frac{c t}{2 x_0}\cdot 2 x_0^3 =
e |{\bf A_g}| 2 x_0^3 \approx 10^{-49} erg\cdot cm^3,$$
$$|C_A| = \frac{1}2 e^2 c t^* x_0 \approx 10^{-49} erg\cdot cm^3.$$
It is easy to show that $|C_V|$ and $|C_A|$ are proportional to $A_g^{-2}$  \cite{33}.  In Ref. \cite{33} the
following expression for probability $W$ of $\beta$-decay is obtained:
$$W \approx C^2_{V,A} E^5 \sim |{\bf A_g}|.$$
 Because really in any point of space the summary potential
$A_\Sigma < |{\bf A_g}|$  and $W \sim A_\Sigma$ are existing, hence one can judge the
possibility of influence of vectorial magnetic potential, for
example, the Earth's one ($\sim 10^8$ CGSE unit), on the $\beta$-decay rate
from its variation.  The experimental investigations on this
subject were carried out in the State Technological University
of St.-Petersbourg. The following equipment and methodology
were used.  The influence of the vectorial potential of the
Earth on electroweak interactions was investigated by way of
counting the number of decays realized of $\beta$-active preparations.
To registrate the decays, the standard detecting units\footnote{
The russian marks of the apparatuses are represented by english letters.}
BDEG-2-23 with the photomultiplier FEU-82 and scintillograph (on NaI) with
dimensions $\oslash 63\times 63$ mm and BDEG-2-20 with the photomultiplier
FEU-49B and scintillograph (on NaI) with dimensions $\oslash 150\times 100 $mm,
were used. The units were fed from the high-voltage sources
BNV-2-95 and VS-22. The signals from FEU entered the analysers
BPA-94M (AI-1024) having the overall number of energy channels
up to 4096. In some measurements, the number of registering
channels was cut down to 1024 for reducing the idle time. All
the equipment was fed on the stabilized voltage, which was
monitored by the digital voltmeter V7-47. During extended
measurements, the analyzer was computer-controlled as well as reading the integrals of events chosen in
separate channels was. As test specimens, the $\beta$-sources GIK-1-4
with cobalt-60 (half-life period $T_{1/2} = 5.3$ years) having
activity $\sim 5\cdot 10^6 $Bc at the instant of measurements and OSGI-2
with cesium-137 ($T_{1/2} = 29.7$ years, activity $\sim 10^6$ Bc), were
used. The characteristic lines of $\gamma$-quanta after $\beta$-decay are,
for Cobalt-60, 1.17 and 1.33 MeV. The preparation of Cesium-137
is more convenient for measurements since it has the single
characteristic line of 0.06 MeV. The de-excitation time in the
scintillograph using NaI is $\sim 250$ ns, nonetheless the ultimate
load of the pulse analyzer AI-4096 without reduction in counting
accuracy is limited by the value of $5\cdot 10^4$ $\gamma$-quanta per second.
Really, the counting rate in the course of measurements was no
more than 16-25 thousands quanta per second. The maximum number
of events recorded in one energy channel is limited by $\sim 65$
thousands. Therefore, in order to prevent overflows, especially
in channels corresponding to characteristic lines, the exposure
accumulation time was forced to be limited with $\sim 400 s$ for
Co-source and $\sim 200 s$ for Cs-source. It has been possible in
this time to registrate $\sim 5-10$ millions quanta, of which $\sim 25-70 \%$
fell on characteristic lines. However taking into account that
the total number of events (photons) registered is proportional
to the number of $\beta$-decays, one may detect variations in decay
rate from their fractional changes.  The principal result of
measurements for the Cs-137 preparation is shown in Fig.1. The
value of oscillation was equal to 25 thousand photons (at the
initial level of $\sim 18.6$ millions per $\sim 16$ min.) which corresponds
to $\sim 6 \sigma$.  The repetition of this effect during the entire cycle
of measurements (Apr.19-23,1994) indicates that it is not
accidental.

\pagebreak

\begin{figure}
\protect\hbox{\psfig{file=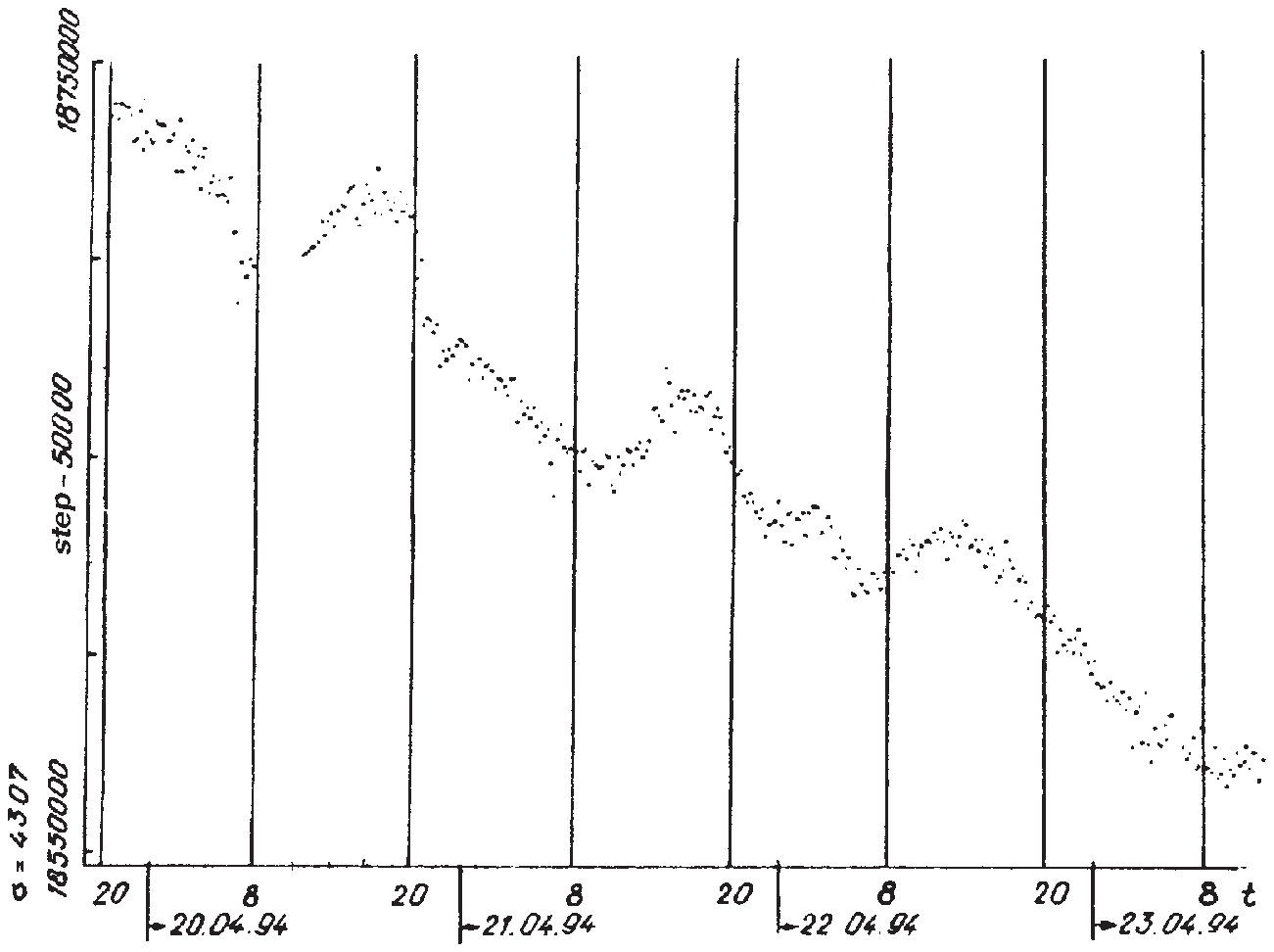,height=16.0cm,width=6.0cm,angle=90}}
\caption{Time variation of Cs-137 $\beta$-decay:
n~-~the number of decays per 16 min,
\hspace*{9cm} t~-~the time of the day (St.-Petersbourg)}
\end{figure}

\end{document}